\documentclass[twocolumn,showpacs,preprintnumbers,amsmath,amssymb]{revtex4}
                                                                                
\usepackage{graphicx}
\usepackage{dcolumn}
\usepackage{subfigure}
\usepackage{bm}
\begin{document}
                                                                                
\preprint{}

\title{Modelling savings behavior of agents in the kinetic exchange models of market}
                                                                                
\author{Anindya S. Chakrabarti}
\affiliation{Indian Statistical Institute, 203 B. T. Road, Kolkata-700108}

\date{\today}

\begin{abstract}
Kinetic exchange models have been successful in explaining the shape of the income/wealth distribution
in the economies. However, such models usually make some ad-hoc assumptions when it comes to determining
the savings factor. Here, we examine a few models in and out of the domain of standard neo-classical economics to
explain the savings behavior of the agents.
A number of new results are derived and the rest conform with those obtained earlier. Connections are established between
the reinforcement choice and strategic choice models with the usual kinetic exchange models.
\end{abstract}

                                                                                
\maketitle

\footnotetext{email-id : {\it aschakrabarti@gmail.com}}

\section{Introduction} 
\noindent The distributions of income
and wealth have long been found to possess some robust and stable
features independent of the specific economic, social and political conditions of the economies.
Traditionally, the economists have preferred to model the left tail and the mode of the distributions of 
the workers' incomes
with a log-normal 
distribution and the heavier right tail with a Pareto distribution.
For a detailed survey of the distributions used to fit the income and wealth data see Ref. \cite{kleiber;03}.
However, there have been several studies recently that argue that
the left tail and the mode of
the distribution fit well with the gamma distribution and the right tail of the distribution follows a power
law \cite{acsybkc;05,acbkc;07,chakrabartis;10}. It has been argued that this feature might be considered to be a natural law
for economics \cite{chakrabartis;10,yako-rosser;09}.

\noindent The Chakraborti-Chakrabarti \cite{CC;00} model (CC model henceforth) can explain the gamma-like 
distribution very well whereas 
the Chatterjee-Chakrabarti-Manna \cite{CCM;03}
model (CCM model henceforth) explains the origin of the power law. Both models fall in 
the category of the kinetic exchange models.
However, the first model assumes a constant savings propensity and the second model assumes an uniformly
distributed savings propensity of the agents. 
It may be noted that the distribution of the savings factor is exogenous in these models. It is not derived from
any optimization on the agent's part or by some other mechanism.

Ref. \cite{chakrabartis;09} considers an exchange economy populated with
agents having a particular type of utility function and derives the CC model in the settings of a
competitive market. Here, we show that the same methodology could be applied to derive the CCM
model and we can explain the savings factor accordingly. 
A further possibility is also studied where the constancy of the savings propensity over time is relaxed.
More specifically, we examine the cases where the savings propensity is dependent
on the current money holding of the corresponding agent. This model, as we shall see, shows self-organization and
in some cases, it gives rise to bimodality in the money distribution.

\noindent It is well known the models of utility maximization has been severely critisized on the grounds of
limitations of computational capability of human beings \cite{bak;97}. 
Hence, to model the savings behavior of the agents,
we study some simple thumb-rules and derive the distributions of savings therefrom and finally the
income/wealth distribution. In particular, we consider the savings propensity as
a strategy variable to the agents. Two cases are explored here. In the first case, the agents take their savings
decisions of their own by reinforcing their choices. In the second case, the agents adopt the winning strategy.
In all cases, we study the final money distributions.

\noindent The plan of this paper is as follows. In section, \ref{sec:CC-CCM} we explain the savings behavior
of the agents using arguments from neoclassical economics. In the next section, we study a self-organizing
market where the savings is a function of the current money holding of the corresponding agent. In section
\ref{sec:onetomany} and \ref{sec:manytoone}, we study the savings behavior of the agents 
where the savings decision follows some very
simple thumb-rules. Then follows a summary and discussion.

\section{Kinetic exchange models in a competitive market}
\label{sec:CC-CCM}
\noindent By competitive market we mean a market with atomistic agents who trade with each other knowing that
their individual actions can not possibly influence the market outcome (that implies there is no {\it strategic}
interaction between the agents). We assume that markets are always
cleared by equating supply and demand; that is the market is completely free of any friction. Below, we elaborate
on the market structure and the behavior of the agents more explicitly.

\subsection{The Chakraborti-Chakrabarti (CC) Model}
\label{sub:CC}

\noindent This model considers homogenous agents characterized by a single savings propensity. We briefly review the
derivation of the model following Ref. \cite{chakrabartis;09}. The structure of the economy
is the following. It is an exchange economy populated with $N$ agents each producing a single
perishable commodity. There is complete specialization in production which means 
none of the agents produce the commodity produced by another. Money is not produced in the economy. All agents are
endowed with a certain amount of money at the very begining of all tradings. Money can be
treated as a non-perishable commodity which facilitates transactions.
All commodities along with money can enter the utility function of any agent as arguments.
These agents care for their future consumptions and hence they care
about their savings in the current period as well. It is natural that with 
successive tradings their money-holding will change with time.
At each time step, two agents are chosen at random to carry out transactions among themselves
competitively. We also assume that 
the preference structure of the agents are time-dependent that is 
the parameters of the utility functions vary over time (Ref. \cite{lux;05,takamoto;02}).
For a detailed discussion on the derivation of the resulting money-transfer equations, see Ref. \cite{chakrabartis;09}.
Below, we provide the formal structure and the solution to the model only.

\noindent Let us assume that at time $t$, agent $i$ and $j$ have been chosen.
Also, assume that agent $i$ produces $Q_i$ amount of commodity $i$ only and agent $j$ produces
$Q_j$ amount of
commodity $j$ only and the amounts of money in their possession at
time $t$ are $m_i(t)$ and $m_j(t)$ respectively (for simplicity, $m_k(0)=1$ for $k$=1,2). 
Notice that the notion of complete specialization in production process
provides the agents with a reason for trading with each other.
Naturally, at each time step
there would be a net transfer
of money from one agent to the other due to trade.
Our focus is on how the amounts money held by the agents change over time due to the repetition of such a trading process.
For notational convenience, we denote $m_k(t+1)$ as $m_k$
and $m_k(t)$ as $M_k$ (for $k=1, 2$).

\noindent Utility functions are defined as follows.
For agent $i$, {$U_i(x_i,x_j,m_i)= x_i^{\alpha_i}x_j^{\alpha_j}m_1^{\lambda}$}
and for agent $j$,
$U_j(y_i,y_j,m_j)=y_i^{\alpha_i}y_j^{\alpha_j}m_j^{\lambda}$ where the
arguments in both of the utility functions are consumption of the first (i.e.,
$x_i$ and $y_i$) and second good (i.e., $x_j$ and $y_j$)
and amount of money in their possession respectively.
For simplicity, we assume that the sum of the powers is normalized to $1$ i.e.,
$\alpha_1+\alpha_2+\lambda=1$.
Let the commodity prices
to be determined in the market be denoted by $p_i$ and $p_j$.
Now, we can define the budget constraints as follows. For agent $i$
the budget constraint is $p_ix_i+p_jx_j+m_i\leq M_i+p_iQ_i$ and
similarly, for agent $j$ the constraint is
$p_iy_i+p_jy_j+m_j \leq M_j+p_jQ_j$.
In this set-up, we get the market clearing
price vector ($\hat p_i, \hat p_j$) as
$\hat p_k=(\alpha_k/\lambda)(M_i+M_j)/Q_k$ for $k=1$, $2$.

\noindent By substituting the demand functions of $x_k$, $y_k$ and $p_k$ for $k=1$, $2$ in the money demand functions,
we get the most important equation of money exchange in this model.
To get the final result, we substitute $\alpha_i/(\alpha_i+\alpha_j)$
by $\epsilon$ to get the money evolution equations as 
\begin{eqnarray}
m_i(t+1) &=& \lambda m_i(t)+\epsilon(1-\lambda)(m_i(t)+m_j(t)) \nonumber\\
m_j(t+1) &=& \lambda m_j(t)+(1-\epsilon)(1-\lambda)(m_i(t)+m_j(t)) \nonumber\\
\label{eqn:CC}
\end{eqnarray}
\noindent where $m_k(t)\equiv M_k$ and $m_k(t+1)\equiv m_k$ (for $k$= $i$, $j$). 
Note that for a fixed value of $\lambda$, if $\alpha_i$ 
is a random variable with uniform distribution over the domain $[0,1-\lambda]$,
then
$\epsilon$ is also uniformly distributed over the domain $[0,1]$.
For the limiting value of $\lambda$ in the utility function
(i.e., $\lambda\rightarrow0$),
we get the money transfer
equation describing the random sharing of money without savings.

{\bf Interpretation of $\lambda$}: Here, it is clearly shown that $\lambda$ in the CC model is nothing but
the power of money
in the utility function of the agents and finally this turns out to be the fraction of money holding that remains
unaffected by the trading action. However, in this form it can not be directly interpreted as the propensity to save.
Below, we try to derive $\lambda$ from an utlity maximization problem while retaining the kinetic exchange structure
and we show that in this slightly alternative formulation, $\lambda$ is indeed the savings propensity as has been
postulated.

\begin{figure}
\begin{center}
\noindent \includegraphics[clip,width= 5cm, angle = 270]
{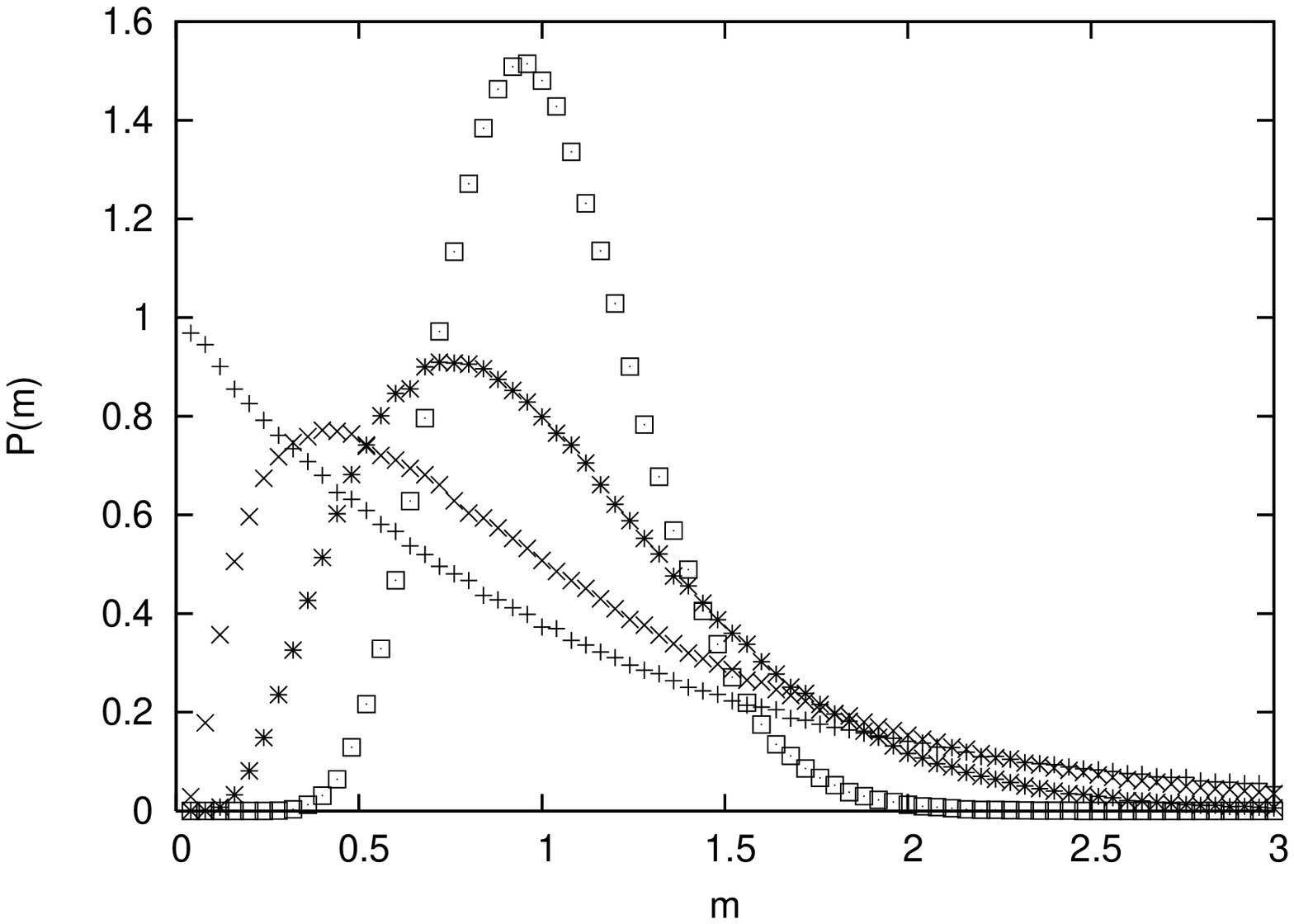}
\caption\protect{The usual CC model: money distribution among the agents. Four cases are shown
above, viz., $\lambda=0$ (+), $\lambda=0.2$ ($\times$),
$\lambda=0.5$ ($\ *$), $\lambda=0.8$ ($\square$).
All simulations are done
for $O(10^6)$ time steps with 100 agents and averaged over $O(10^3)$ time steps.

}
\label{fig:CC}
\end{center}
\end{figure}

\subsection{The Chatterjee-Chakrabarti-Manna (CCM) Model}
\label{sub:CCM}

\noindent As is clear from above, in the CC model the savings decision, the market 
clearence, the prices are all determined at
the same instant. But the savings decision is usually made in separation. More specifically, we can model the savings
decision and the market clearence distinctly. The CCM model takes into account the heterogeneity of the agents. 
In particular, it assumes that it is the
savings propensity of the agents which differs from each other. To derive the same, we assume that the agents
take decisions in two steps. First, they decide how much to save and in the second step, they go to the market with
the rest of the money and take the trading decisions.

\noindent Formally, we can analyze a typical agent's behavior at any time step $t$ in the 
following two steps.
\begin{enumerate}
\item[(i)] Each agent's problem is to make the decision regarding how much to save.
For simplicity, we assume that the utility function
is of Cobb-Douglas type. Briefly, at time $t$ the $i-$th agent's problem is to
maximize $U(f_t,c_t)=f_t^{\lambda_i} c_t^{(1-\lambda_i)}$
subject to $f_t/(1+r)+c_t=m(t) $
where $f$ is the amount of money kept for future consumption, $c$ is
the amount of money to be used for current consumption, $m(t)$ is the
amount of money holding at time $t$ and $r$ is the interest rate
prevailing in the market which can be assumed to be zero in a conservative framework. 
This is a standard utility maximization
problem and solving it by Lagrange multiplier,
we get the optimal allocation for the $i-$th agent as
$f_t^*=\lambda_i m(t)$ and $c_t^*=(1-\lambda_i) m(t)$.
Clearly, this decision is independent of what other agents are doing.
So now the agents will go to the market with $(1-\lambda_i) m(t)$ only.

\item[(ii)]  Now that each agent has made the savings decision, they can engage in competitive trade with each other
in the fashion descibed in subsection \ref{sub:CC} with $\lambda \rightarrow$ 0 (but $\lambda \ne 0$; it is a
mathematical requirement). Note that the amount of money
used by the $i$-th agent is $c_t^*=(1-\lambda_i) m(t)$ only.

The resultant asset exchange equations are those given by the CCM model \cite{CCM;03}.
\end{enumerate}
\begin{eqnarray}
m_i(t+1) &=& \lambda_i m_1(t)+\epsilon[(1-\lambda_i)m_i(t)+ \nonumber\\
~~~~~~(1-\lambda_j)m_j(t)] \nonumber\\
m_j(t+1) &=& \lambda_j m_2(t)+(1-\epsilon)[(1-\lambda_i)m_i(t)+ \nonumber\\
~~~~~~(1-\lambda_j)m_j(t)] \nonumber\\
\label{eqn:CCM}
\end{eqnarray}

\begin{figure}
\begin{center}
\noindent \includegraphics[clip,width= 5cm, angle = 270]
{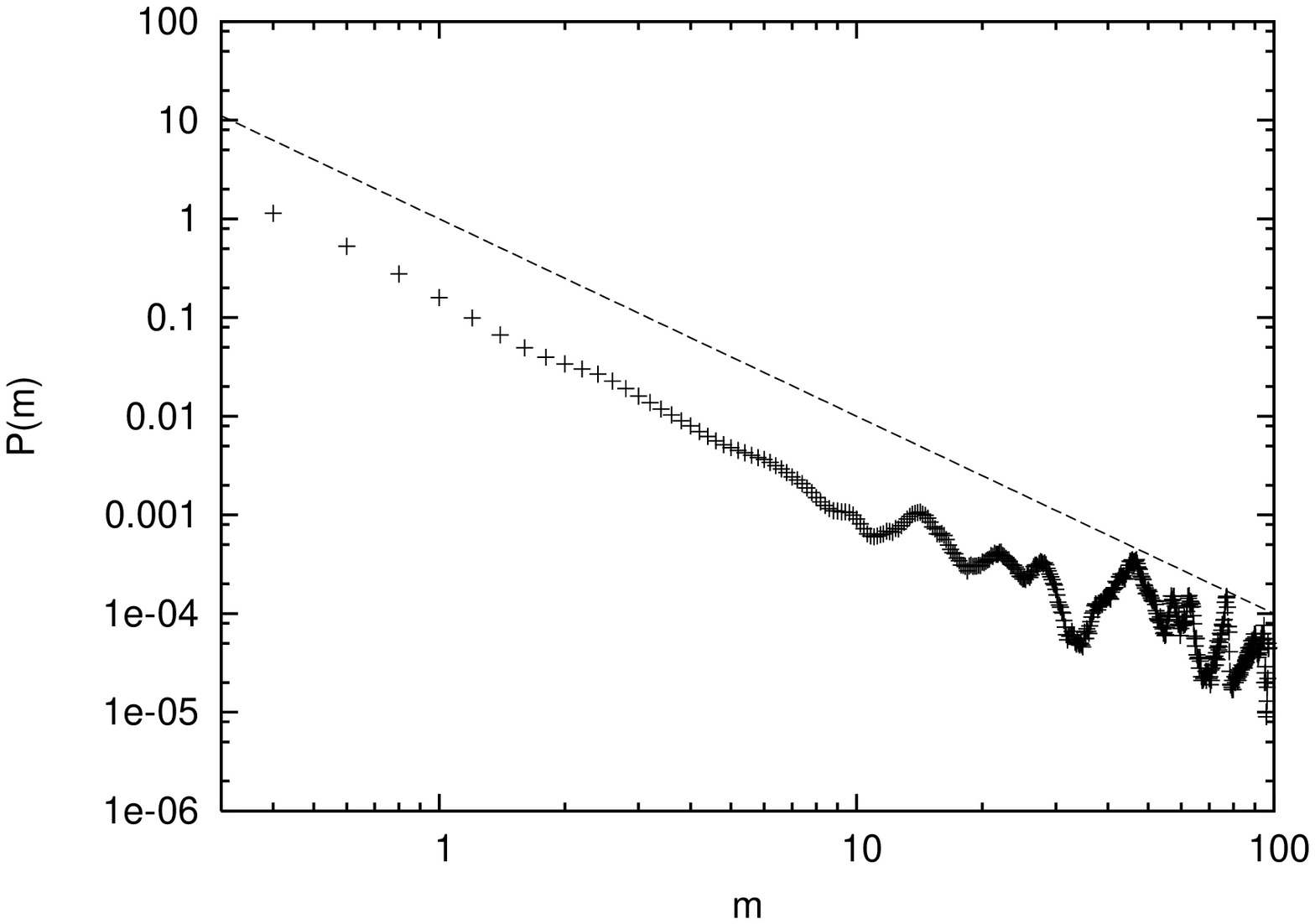}
\caption\protect{The usual CCM model: money distribution among the agents.
All simulations are done
for $O(10^6)$ time steps with 100 agents and averaged over $O(10^3)$ time steps. The straight line
is a guide for a power law with slope -2.

}
\label{fig:CCM}
\end{center}
\end{figure}

{\bf Interpretation of $\lambda_i$}: First we recall the solution of the savings decision which is
$f_t^*=\lambda_i m(t)$. Note that it implies
$$
\lambda_i=\frac{f_t^*}{m(t)}
$$

\noindent that is $\lambda_i$ is nothing but the proportion of money kept for future usage to the current money holding
and this is by definition the savings propensity.

\section{$\lambda$ as a function of money: a self-organizing economy}

\label{sec:self-organization}

\noindent A distinct possibility is that the savings propensity is a function of money-holding itself. To examine that
case, we need $\lambda$:[0,$\infty$)$\rightarrow$[0,1]. However, there are two possibilities. The savings propensity
can be an increasing or decreasing function of money-holding. The simplest forms that we may assume are the following.
\\
\begin{enumerate}
\item[(i)] $\lambda_t=c_1e^{-(c_2m(t))}$ with $c_1$ $<$1: Savings propensity is a decreasing function of money holding. The
restriction on $c_1$ does not allow any agent to have savings propensity equals to 1.
The system shows self-organization and assumes a stable probability density function in the steady state 
(See Fig. \ref{fig:sav-exp-neg} where, for purpose of illustration, $c_1$ has been kept constant at $0.95$).
It is seen numerically that as $c_2$ increases the distribution converges to an exponential density function. In the 
other extreme, it tends to the CC model with $\lambda$ = $c_1$. 
For moderate values of $c_2$, the distribution resembles gamma function.
For other values of $c_1$ also, the system behaves similarly. Note that since $c_1$ is the maximum possible savings
propensity, for a very low value of it, the system becomes indistinguishable from an exponential distribution.
While it seems counter-intuitive that savings propensity falls with the money holding, this might in fact be possible
since poorer people can not take any chance to gamble whereas richer people can.

\item[(ii)] $\lambda_t=c_1(1-e^{-(c_2m(t))})$ with $c_1 <$ 1: Savings propensity 
is an increasing function of money holding. The economy
again organizes itself and the distribution of money becomes stable over time. 
However, there is something more. It is seen numerically that bimodality may
apper spontaneously in the density function of money. See Fig. \ref{fig:sav-exp-pos}. Ref. 
\cite{kargupta;06}
discusses such bimodal distribution of wealth (or money). There
a mixture of the agents was used where two classes of agents were characterized by
two different and widely separated (but fixed!) savings propensities. Such a segregated population gave
rise to bimodal distributions. However, such segregations are exogenous since the $\lambda$s are given from
outside the system. Here, however, we have a new model in which savings decision is completely endogenous and the
economy organizes itself in such a way that it gives rise to a class of bimodal distributions.
It may be noted that bimodality in the income/wealth distribution has actually been observed in many cases 
(see e.g., \cite{sala-i-martin}).

\begin{figure}
\begin{center}
\noindent \includegraphics[clip,width= 5cm, angle = 270]
{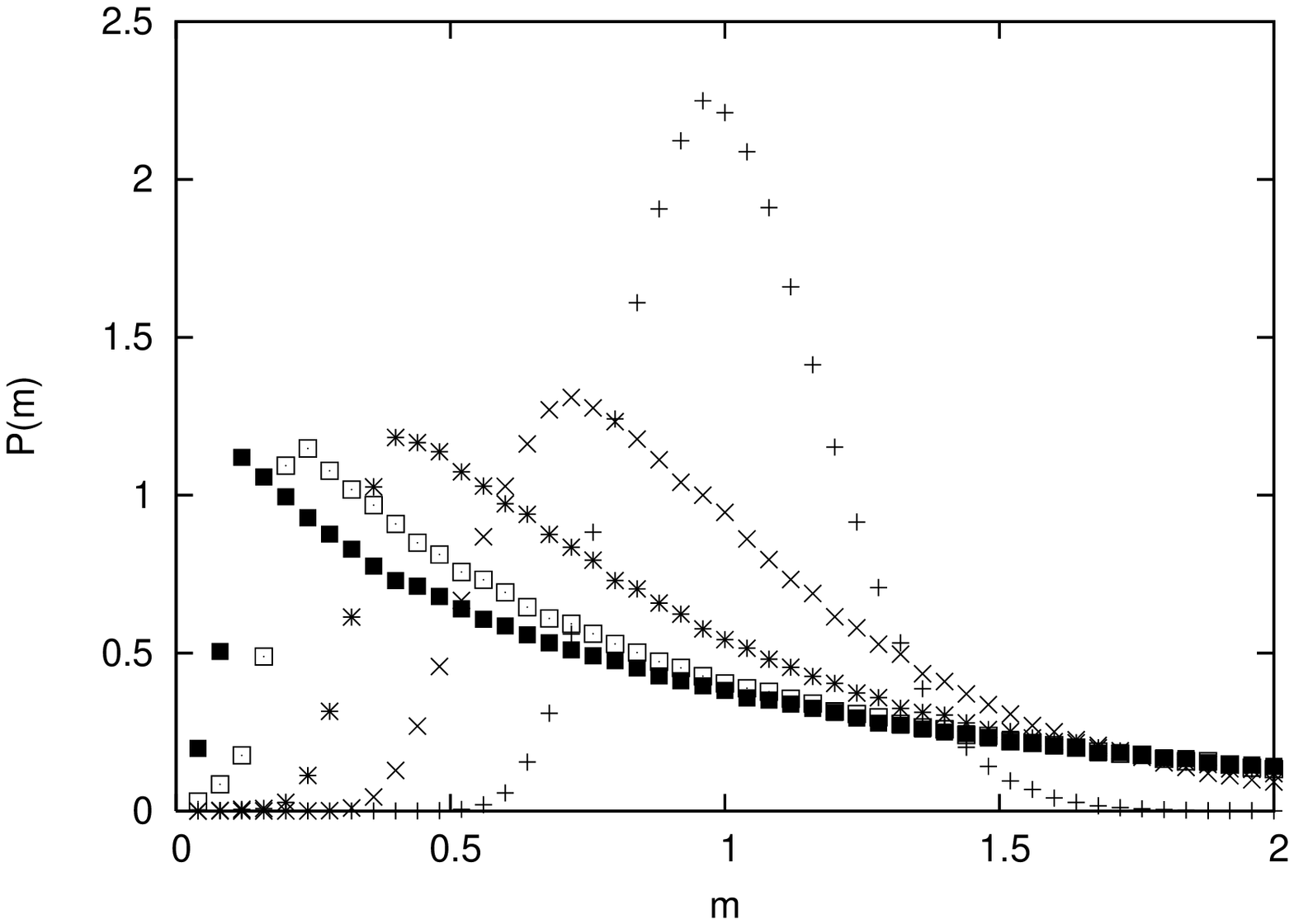}
\caption\protect{Savings propensity is negatively related to the level of money.
All simulations are done
for $O(10^5)$ time steps with 100 agents and averaged over $O(10^3)$ time steps.
Here, $c_1$ has been kept constant at 0.95 and $c_2$ has been changed. The plots include
$c_2$= 0.1 (+), 0.5 ($\times$), 1 ($\ *$), 2 ($\square$), 4 ($\blacksquare$). It is seen that as $c_2$ increases,
the distribution becomes more and more
skewed finally converging to an exponential density function.
}
\label{fig:sav-exp-neg}
\end{center}
\end{figure}

\begin{figure}
\begin{center}
\noindent \includegraphics[clip,width= 5cm, angle = 270]
{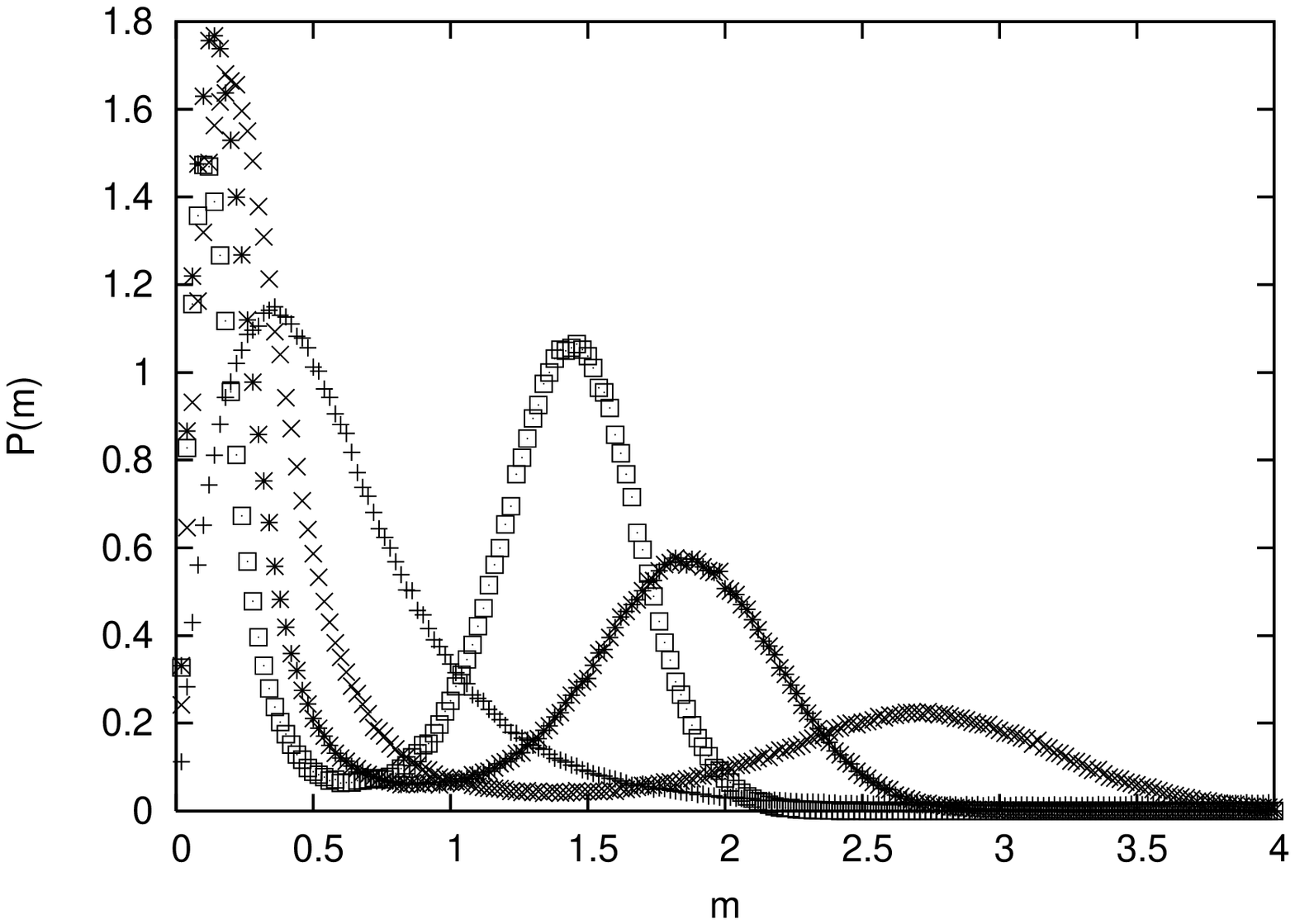}
\caption\protect{Savings propensity is positively related to the level of money.
All simulations are done
for $O(10^6)$ time steps with 100 agents and averaged over $O(10^3)$ time steps.
Here, $c_1$=0.95 and the curves are plotted for $c_2$= 1 (+), 2 ($\times$), 3 ($\ *$) and 4 ($\square$). Bimodality is
clearly seen in the distribution of money.
}
\label{fig:sav-exp-pos}
\end{center}
\end{figure}

\noindent It is seen numerically that bimodality appears for $c_1 \ge 0.92 $ and $c_2 \ge 1$. 
Another interesting feature of this model is that keeping $c_1$ constant as $c_2$ increases, 
the monomodal distribution breaks into
a bimodal distribution which again becomes a monomodal distribution for even larger values of $c_2$. For example,
consider Fig.
\ref{fig:sav-exp-pos} in which the maximum value of $c_2$ considered is 4. But as $c_2$ increases further, the
distribution again becomes monomodal.
However, it should
also be mentioned that if 
$c_1$ is too large (for example, if $c_1\ge$ 0.97), then the system produces some
strange-looking bimodal distributions.

\end{enumerate}

\section{`Irrational' decision making}
\label{sec:irrational}
\noindent The standard economic paradigm of market clearence via utility maximization has been criticised on the
grounds of limitations of computational capability of human beings \cite{bak;97}. The main challenge is to
derive the homogeneity in savings behavior of the agents from a very simple thumb-rule such that the
final distribution of income/wealth looks realistic. A few realistic components of decision-making are noted 
in Ref. \cite{flache;02, pemantle;07}
\begin{enumerate}
\item[(i)] Players develop prefernces for choices associated with better outcomes even though the association
may be coincident, causally spurious, or superstitious.
\item[(ii)] Decisions are driven by the two simultaneous and distinct mechanisms of reward and punishment, which are
known to operate ubiquitously in humans.
\item[(iii)] Satisficing or persisting in a strategy that yields a positive but not optimal outcome, is common and
indicates a mechanism of reinforcement rather than optimization.
\end{enumerate}

\noindent Of particular interest is item (iii) which goes directly against the derivations stated above (see Section
\ref{sec:CC-CCM}).

\section{From one to many ...}
\label{sec:onetomany}
\noindent To model the savings behavior of the agents, we now make the following assumptions.

\begin{enumerate}
\item[(i)] The agents do not perform static optimization.
\item[(ii)] There is reinforcement in their decision-making process.
\item[(iii)] The agents look for better payoffs. But eventually each of them converges to a 
single and simple strategy or thumb rule.
\end{enumerate}

\begin{figure}
\begin{center}
\noindent \includegraphics[clip,width= 5cm, angle = 270]
{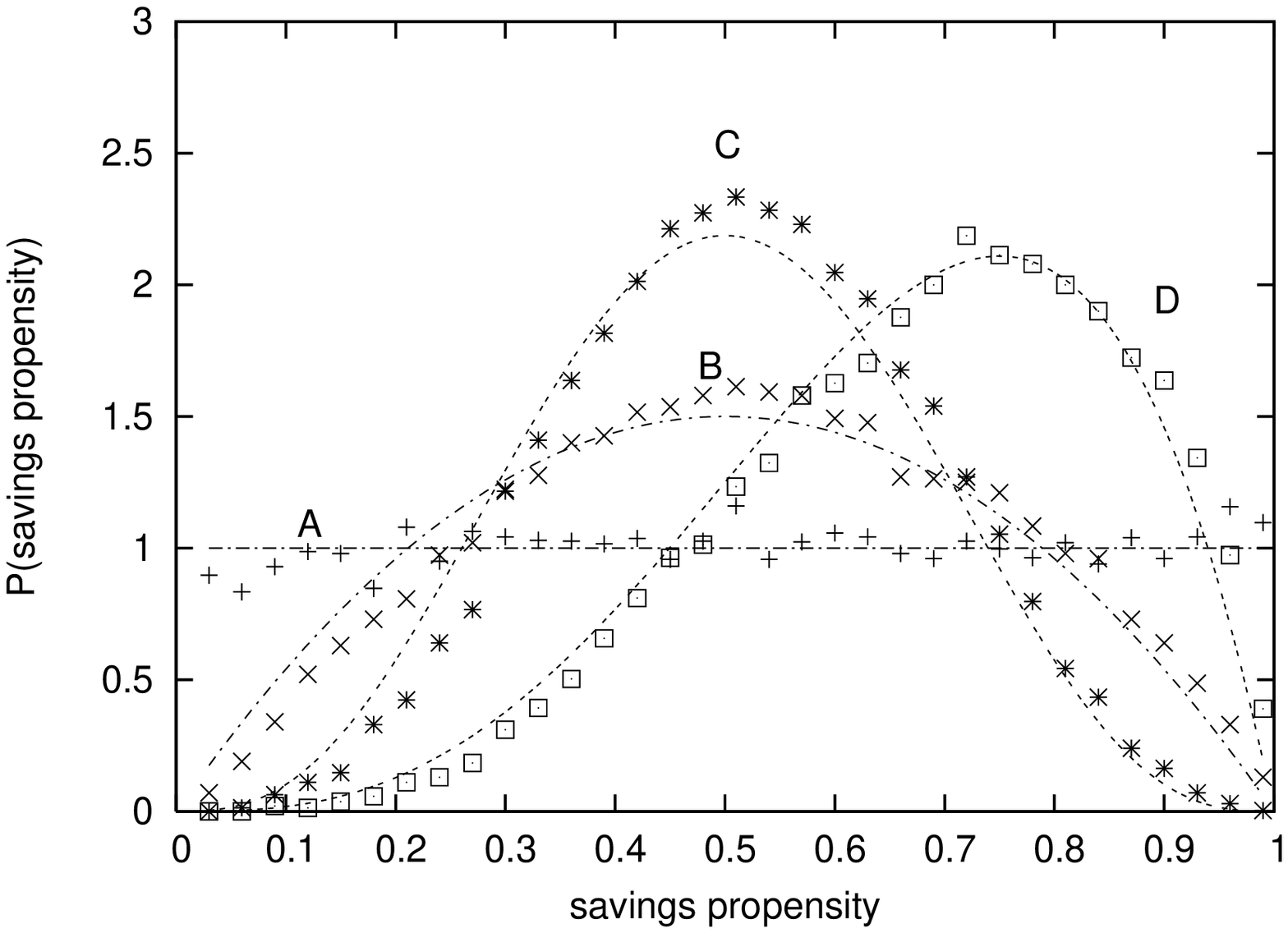}
\caption\protect{The savings distribution by reinforcement. Four cases are shown. Case A: a=1, b=1. Case B: a=2, b=2.
Case C: a=4, b=4. Case D: a=4, b=2. The solid and the dotted lines show the theoretical 
results (the $\beta(a,b)$ distributions).
}
\label{fig:sav-dist-polya}
\end{center}
\end{figure}

\noindent To incorporate the three above-mentioned assumptions, we model the agents' saving behavior by Polya's urn
process \cite{pemantle;07}. The model is as follows. Consider the $i-$th agent. The choice 
is binary, he can take any of the 
two decisions, to consume $(c)$ or
to save $(s)$. His strategies ($c$ and $s$) such that $c,s=0,1$ and $c+s=1$. At each instant, he chooses the values
for $c$ and $s$.
Define $C_t$ ($S_t$) as the number of times $c$ ($s$) has been assigned a value of unity in $t$ time periods.
The savings propensity at time $t$ (that is $\lambda_t$) is defined as the ratio of $S_t$ to $(S_t+C_t)$.
We can assume that initially $S_0=a$ and $C_0=b$. The reinforcement mechanism is incorporated by assuming that
the probability of choosing $s$ = 1 at any time $t+1$, is simply $\lambda_t$.
Basically, it is the Polya's urn model and the famous result that follows from it is 
the following (Ref. \cite{pemantle;07}).
{\it The random variable $\lambda_t$ converge almost surely to a limit $\lambda$. The distribution of $\lambda$ is
$\beta(a,b)$.} (See Fig. \ref{fig:sav-dist-polya} ).

\begin{figure}
\begin{center}
\noindent \includegraphics[clip,width= 5cm, angle = 270]
{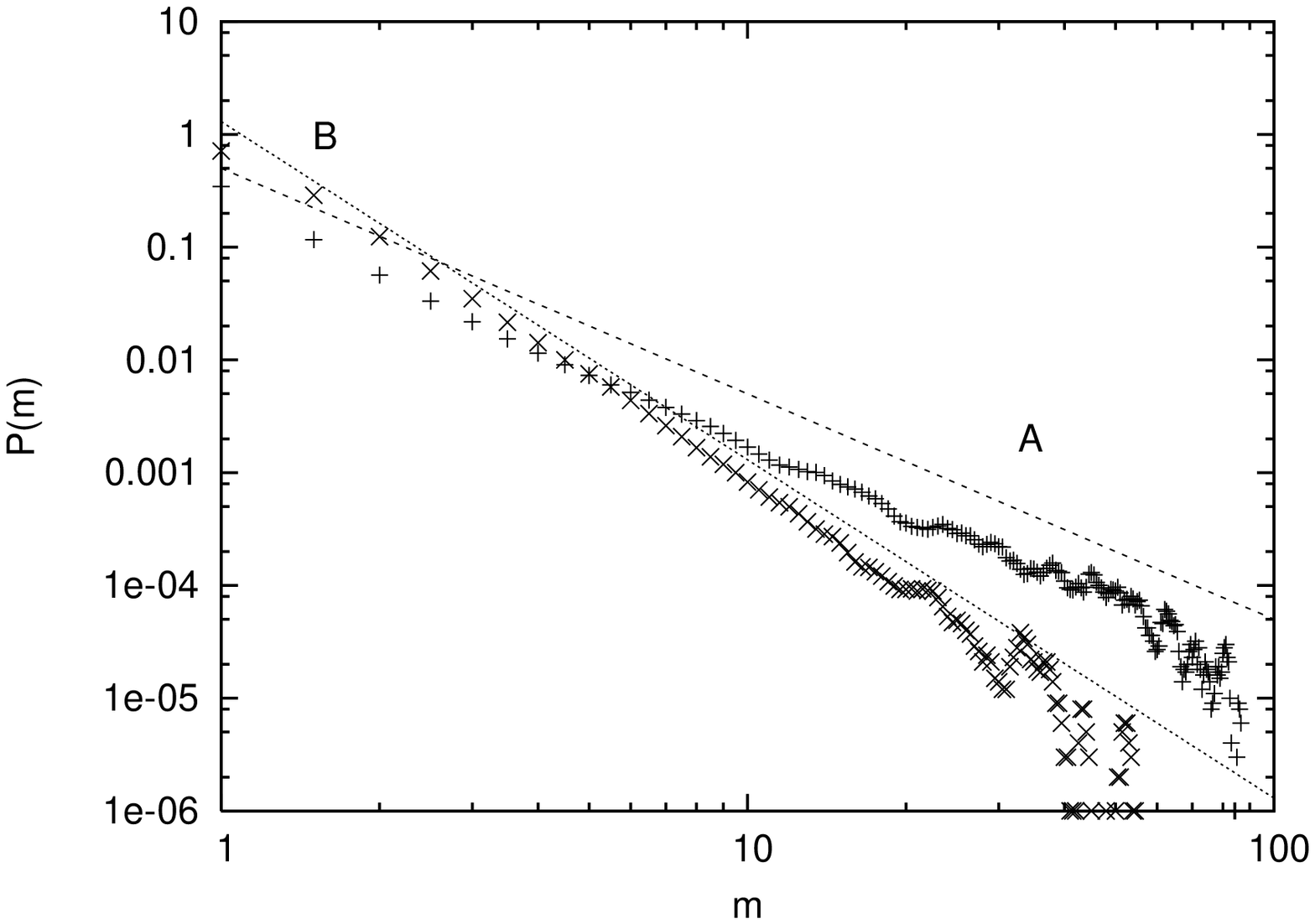}
\caption\protect{The steady state money distributions for two cases:- ($A$) $a$=1, $b$=1 and ($B$) $a$=4, $b$=2. 
Case ($A$) produces
the same distribution of money as the CCM model (slope -2 in log-log plot) whereas in case ($B$) we get a
distribution with slope -3 in the log-log plot. The lines drawn have slopes -2 and -3 respectively.
All simulations are done
for $O(10^6)$ time steps with 100 agents and averaged over $O(10^3)$ time steps.

}
\label{fig:income-dist-polya1}
\end{center}
\end{figure}

\begin{figure}
\begin{center}
\noindent \includegraphics[clip,width= 5cm, angle = 270]
{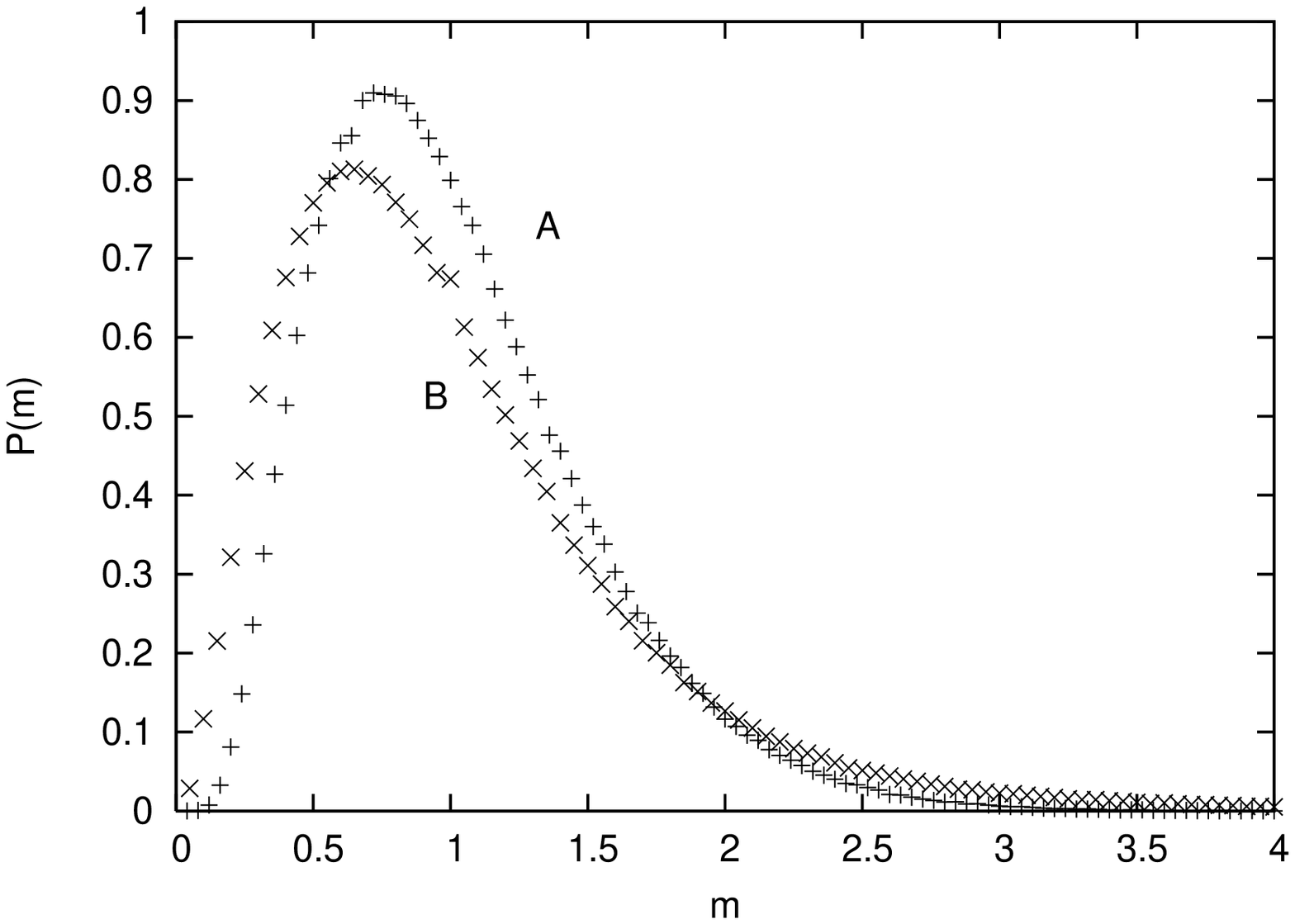}
\caption\protect{The steady state money distributions for two cases:- ($A$) $a=b$; $a$, $b$$\rightarrow$ $\infty$ 
and ($B$) $a$=4, $b$=4. Note that case ($A$) is identical to CC model with $\lambda$ = 0.5.
All simulations are done
for $O(10^5)$ time steps with 100 agents and averaged over $O(10^2)$ time steps.

}
\label{fig:income-dist-polya2}
\end{center}
\end{figure}

\subsection{Two limits}
\begin{enumerate}
\item[(i)] ($a$=$b$=1): From the above result, $\lambda_t$ converges to $\lambda$ 
where $\lambda\sim$ uni[0,1]. The resultant
distribution of money follows a power law. This is the basic CCM model. See Fig. \ref{fig:income-dist-polya1}.

\item[(ii)] ($a$=$b$; $a,b$$\rightarrow$ $\infty$): $\lambda_t$ converges to $\lambda$ where $\lambda$ is a delta function
at 0.5. This corresponds to a special case of the CC model where $\lambda=0.5$. See Fig. \ref{fig:income-dist-polya2}.
\end{enumerate}

\subsection{For moderate values of $a$ and $b$}

Clearly, (1$<$$a$, $b$$<$$\infty$): This model gives the gamma-like part as well as the Pareto tail of the income/money
distribution for different values of $a$ and $b$. For example, we show the results of two cases. 

\begin{enumerate}
\item[(i)] ($a$=4, $b$=2): The resulting distribution of savings propensity is clearly $\beta(4,2)$. The distribution of
money in the steady state follows a power law with a slope -3 in the log-log plot.
See Fig. \ref{fig:income-dist-polya1}.

\item[(ii)] ($a$=4, $b$=4): The resulting distribution of savings propensity is $\beta(4,4)$. The distribution of money in
the steady state is gamma function-like.
See Fig. \ref{fig:income-dist-polya2}.

\end{enumerate}

\section{From many to one ...}
\label{sec:manytoone}
\noindent In Section \ref{sec:onetomany}, we have discussed how one can derive a set of distributed savings propensities
starting from
a unique value. Here, we discuss the reverse side of the same coin. We shall show that the agents with
different savings propensities, can converge to a single value over time.

\noindent To model this situation, we assume that the agents treat savings propensity as a strategy which evolves over time.
A very simple rule of evolution is the following. The winner in any trade retains his strategy whereas the loser adopts
the winner's strategy. Note that by winning in a trading action, we simply imply that the agent who 
gets the lion's share in that particular trading is the winner. Since it is a relative term, (by refering to 
Eqn.\ref{eqn:CCM}) 
winning is determined by the value of the stochastic term $\epsilon$. If $\epsilon \ge 0.5$, then the $i$-th agent wins (in
Eqn. \ref{eqn:CCM}) else the $j$-th agent wins. Note that the most important support of this type of strategy evolution
comes from the third observation by Flache and Macy \cite{flache;02} noted in Section \ref{sec:irrational}.

\noindent Let us assume that the possible saving propensities are finite and denoted by
$\lambda_1$, $\lambda_2$, ..., $\lambda_k$ etc. Also, let us denote the number of agents with $\lambda_i$ savings
propensity at time $t$ by $n_i(t)$ (for $i$= 1, ..., k). 
At each time-period two of the agents are randomly selected and they trade according to
Eqn. \ref{eqn:CCM} and then the loser adopts the winner's savings propensity. This process is repeated untill the sytem
reaches a steady state {\it in terms of savings propensities}. After the whole system becomes steady with the agents
with a fixed saving propensity, the system is allowed to evolve further to reach a steady state in terms of money.

\subsection{\bf Convergence in savings propensity} Let 

\begin{equation}
\sum_{i=1}^{k}n_i(0)=N,
\label{sum of agents}
\end{equation}

\begin{figure}
\begin{center}
\noindent \includegraphics[clip,width= 5cm, angle = 270]
{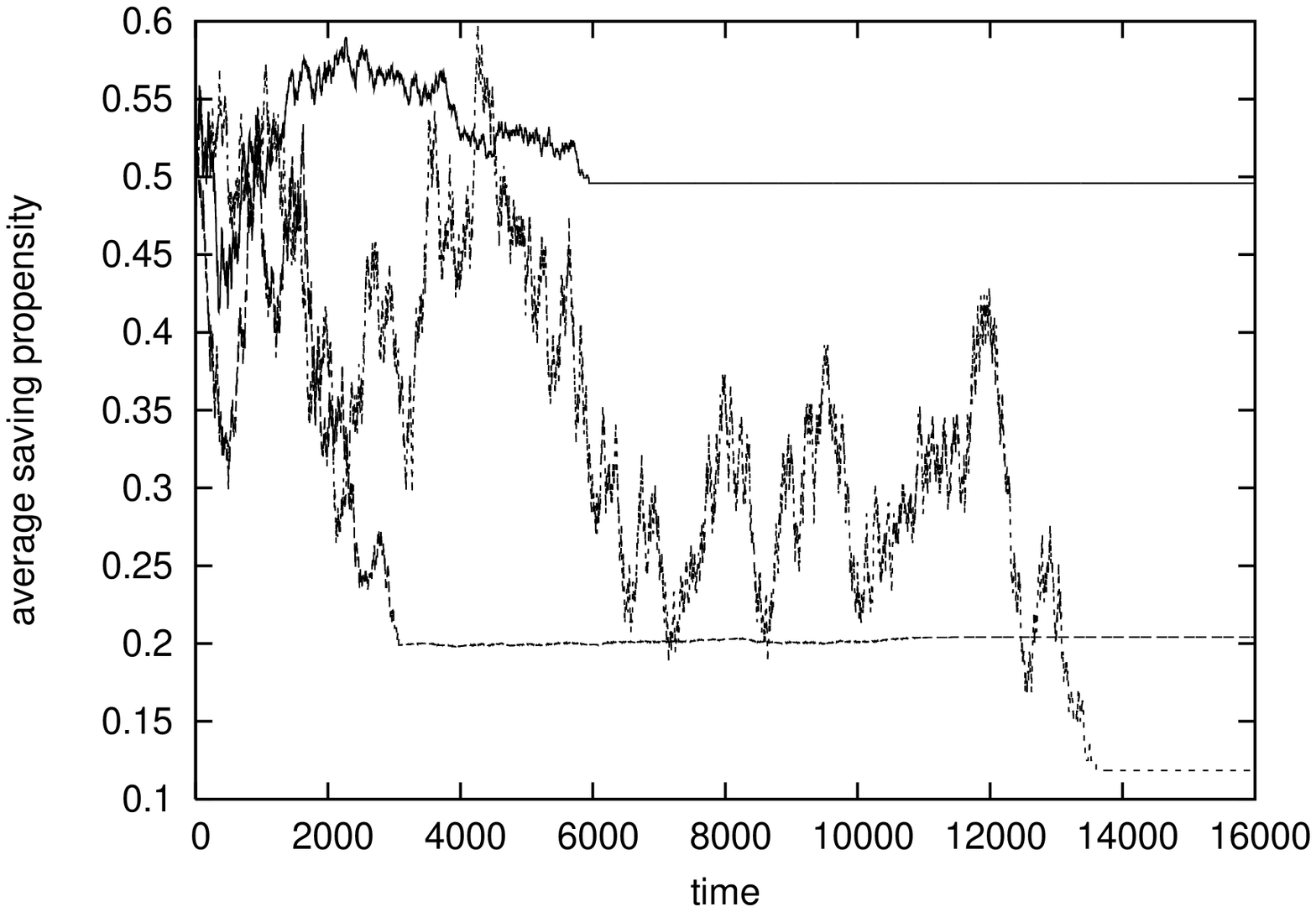}
\caption\protect{The convergence in savings propensity. As examples, three cases are shown.
To study the phenomena of convergence in the savings propensities
we focus on the fluctuation of the average savings propensities.
Initially all agents are assigned with uniformly distributed
savings propensities. Hence the average savings propensity is initially very close to half. But over time it
evolves to reach a steady state where it maintains a fixed value indicating that all the agents have the same 
savings propensity. Thereafter the system behaves like the usual CC model.
}
\label{fig:sav-conv}
\end{center}
\end{figure}

\noindent as the total number of agents remains fixed over time (recall that $n_i(t)$ has been defined above as the 
number of agents with a particular savings propensity $\lambda_i$). 
The agents only shift from one savings propoensity to
another over time. Note that at each (trading) time point, the number of agents with a particular savings propensity
rises or falls by unity with equal probability (i.e. depending on whether $\epsilon \ge 0.5$ or not) or it remains unchanged
if its agents are not selected to trade. To put it formally, let us assume that the two agents selected two trade have
savings propensities $\lambda_i$ and $\lambda_j$ respectively. Then

\begin{equation}
n_i(t+1)=n_i(t)\pm 1  \hspace{.5 cm} \mbox{with equal probability}
\end{equation}
\begin{equation}
n_i(t+1)+n_j(t+1)=n_i(t)+n_j(t)
\end{equation}

\begin{equation}
n_k(t+1)=n_k(t) \hspace{.5 cm} \mbox{for all k$\ne$ $i$,$j$}. 
\end{equation}

\noindent Hence, the number of agents with a particular savings propensity performs a random walk of unit step and also note
that the walk is bounded below since $n_i\ge0$ for all $i$ and also above by Eqn. \ref{sum of agents}. 
In fact, the formulation is
akin to the {\it n-players ruin problem} \cite{swan} where the random walk occurs on a simplex given by Eqn. 
\ref{sum of agents}. 
The general {\it n-players ruin problem} is very difficult to solve analytically for $k\ge 4$.
Ref. \cite{swan} presents a matrix-theoretic approach to the problem which reduces the complexity of the computation.
However, we are not interested in finding the exact solutions to the problem. We simply note that the given enough time
the system will ultimately evolve to a state where there is only one savings propensity and this is not unique. It
can be any of the initial $\lambda_i$s ($i$ = 1, 2,,, k). See Fig. \ref{fig:sav-conv}.

\subsection{\bf Steady state money distribution} Once all the agents have a single savings propensity, the system then
behaves like the CC model (see Fig. \ref{fig:CC}).

\section{Summary and Discussion} 
\label{summary}
\noindent The kinetic exchange models have been very successful in explaining the origin of the
gamma function-like distribution and
the power law in the income/wealth distribution. However, these models use the notion of savings extensively
on an ad-hoc basis without offering much theoretical understanding of it. The aim of the present paper is
to provide support to the kinetic exchange models by deriving and explaining them from standard neo-classical
economics paradigm and the not-so-standard models of reinforcement learning and strategic selection.

\noindent Ref. \cite{chakrabartis;09} provides a microeconomic basis for the kinetic exchange models with homogenous
agents. Here, we extend that model to explain the heterogenous exchange models where the agents have 
different savings propensities (Sec. \ref{sec:CC-CCM}). 
A further possibility is investigated in Sec. \ref{sec:self-organization} where the savings propensity of an agent is
dependent on the money holding of that particular agent and hence it changes over time (as the money holding changes).
It is shown that even in that case, the economy organizes itself in such a way that the distribution of
money becomes stable over time. In some cases, the distribution produces bimodality. Bimodal income/wealth
distributions have indeed been seen in many countries (see e.g., Ref. \cite{sala-i-martin}). 

\noindent However, it is also noted in Sec. \ref{sec:irrational} that the
market clearing, competitive models used extensively in the economics literature has been 
criticised on the grounds of limitations
of computational capability of human beings (see e.g., \cite{bak;97}). 
So we try to explain the kinetic exchange models assuming that
the agents follow some simple rules of thumb. It is shown that the mechanism of reinforcing one's own choice leads to the
CCM model \cite{CCM;03} (Sec. \ref{sec:onetomany}). The basic result regarding the distribution of the fixed points 
follows from the famous {\it Polya's Urn} problem (Ref. \cite{pemantle;07}). Next, we show in Sec. \ref{sec:manytoone} 
that the agents
following a simple rule of thumb of selecting the best strategy leads to the CC model (Ref. \cite{CC;00}). The game of
strategy selection reduces to the generalized {\it Gambler's Ruin} problem or the {\it N-player Ruin} problem 
(Ref. \cite{swan}).

\begin{acknowledgements}
The author is grateful to Bikas K. Chakrabarti and Arnab Chatterjee for some useful suggestions.

\end{acknowledgements}

\end{document}